\documentclass{aa} 
\usepackage{graphicx}
\usepackage{txfonts}
\usepackage{wasysym}
\usepackage{natbib}

\bibpunct{(}{)}{;}{a}{}{,}
\begin{document}
   \title{Molecular Disks in Radio Galaxies}
   \subtitle{The pathway to ALMA}

   \author{I. Prandoni\inst{1}\fnmsep\thanks{email:prandoni@ira.inaf.it}
          \and 
          R.A. Laing\inst{2}
          \and
          H.R. de Ruiter\inst{1}
          \and
          P. Parma\inst{1} 
          }

   \institute{INAF - Istituto di Radioastronomia, Bologna, Italy\\
         \and
             European Southern Observatory, Garching b. Munch\"en, Germany
             }

   \date{Received 22 July 2010 / Accepted 30 August 2010}

 
  \abstract
   {It has recently been proposed that the jets of low-luminosity radio galaxies are powered 
by direct accretion of the hot phase of the IGM onto the central black 
hole. 
Cold gas remains a plausible alternative fuel supply, however. 
The most compelling evidence that cold gas plays a role in fueling radio galaxies is that 
dust is detected more commonly and/or in larger quantities in (elliptical) radio galaxies compared
 with radio-quiet elliptical galaxies.  On the other hand, only 
\rm
small numbers of radio galaxies have yet been detected in CO (and 
even fewer imaged), and whether or not all radio galaxies have enough
cold gas to fuel their jets remains an open question. If so, then  
the dynamics of the cold gas in the nuclei of radio galaxies may provide 
important clues to the fuelling mechanism. }
   {The only instrument capable of imaging the molecular component on 
scales relevant to the accretion process is ALMA, but very little is yet 
known about CO in southern radio galaxies. Our aim is to measure the CO 
content in a complete volume-limited sample of southern radio galaxies, 
in order to create a well-defined list of nearby targets to be imaged in 
the near future with ALMA.}
   {APEX\thanks{This publication is based on data acquired with the Atacama Pathfinder Experiment (APEX). APEX is a collaboration between the Max-Planck-Institut fur Radioastronomie, the European Southern Observatory, and the Onsala Space Observatory.} has 
recently been equipped with a receiver (APEX-1) able to observe the 230 GHz waveband. This allows us to search for CO($2-1$) line emission in our target galaxies. }
   {Here we present the results for our first three southern targets, proposed for APEX-1 spectroscopy during science verification: 
   \object{NGC\,3557}, 
\object{IC\,4296} and \object{NGC\,1399}. The experiment was successful with two targets detected, and  possible indications for a double-horned CO line profile, consistent with ordered rotation.
These early results are encouraging, demonstrating that APEX can efficiently detect CO in nearby radio galaxies. We therefore plan to use 
APEX to obtain CO spectroscopy for all our southern targets.  }
   {}


   \maketitle

\section{Introduction}\label{sec-intro}
It is generally thought (see e.g. Heckman et al.~\cite{Hetal86}; Baum et al.~\cite{Betal92}) 
that  powerful radio galaxies (typically Fanaroff-Riley type II or FR\,II,
Fanaroff \& Riley~\cite{FR74}) are triggered by mergers or collisions with gas-rich
galaxies. 
During the merger event, part of the gas can be transported to the central region of the merging 
system, where it may trigger a starburst, and, if it loses sufficient angular momentum, feed 
the AGN. The radio jets may then be powered either from the gravitational energy released
during the accretion process, or by electromagnetic extraction of the spin energy of the black
hole (Blandford \& Znajek~\cite{BZ}), which also requires an accretion flow to anchor
a magnetic field.

On the other hand, it has been suggested recently that accretion in low luminosity (or 
more generally low excitation) radio galaxies, often, but not exclusively, Fanaroff-Riley type I or 
FR\,I, may occur directly from the hot phase of the interstellar medium (Allen et al.~\cite{Alletal06}; Evans et al.~\cite{Evetal08}).
Hot ($T > 10^7$\,K) plasma is observed to be ubiquitous in radio-galaxy
nuclei, and Evans et al.\ (\cite{Evetal08}) suggest that the gravitational energy released
if accretion occurs at the 
Bondi rate may be sufficient to power the jets in  FR\,I radio galaxies. It is unclear whether this rate is relevant, however, since 
it assumes spherically symmetric accretion and takes no account of the 
need for angular momentum loss. The accretion rate in the vicinity of the black
hole cannot be estimated directly and may be much lower. In any case, McNamara et al.\ (\cite{McN2010}) find that accretion at the Bondi 
rate is generally unable to power the jets in brightest cluster galaxies and argue that
electromagnetic extraction of the black hole spin energy is required.

An alternative fuel source is cold gas, which is often present in large 
amounts, and is detectable via emission and absorption by dust, HI and 
CO. This cold gas quite often comes in the form of dusty disks, as 
can be clearly seen in
high resolution HST images (Capetti et al.~\cite{Cetal00}; Verdoes-Kleijn~\cite{VK99}; de 
Ruiter et al.~\cite{deRetal02}). Dust is observed in 53\% of the B2 sample of nearby radio galaxies 
(mostly FR\,I) and the dust mass is correlated with radio power 
(de Ruiter et al.~\cite{deRetal02}). 
Similar results have been found for a sample of nearby  3CR radio galaxies. 
At least 30\% of the sources show evidence for dust absorption. Dust in FR\,I radio galaxies is generally
situated in well defined disks on small ($<2.5$ kpc) scales, and the radio source axis tends to lie nearly perpendicular to the dust
disk. Very interestingly FR\,I galaxies have derived dust masses that are typically larger than the dust masses found in a matched sample of radio-quiet ellipticals (de Koff et al.~\cite{deK00}). 
If the mass of dust and cold gas are also correlated, as seems plausible, then this association argues that accretion of cold gas may indeed power the radio jets.

If accretion of (initially) cold gas is the dominant fuelling mechanism, 
then we should be able to detect the gas in {\sl every} low-excitation 
radio galaxy. In normal, extended, FR\,I sources very little HI gas is found
close to the nucleus  
(certainly no more than in early-type galaxies of similar mass selected without
reference to radio emission); in most cases only upper limits M$_{\rm HI} \la
10^8 M_{\odot}$ are available (Emonts et
al.~\cite{Emetal10}). On the other hand, we know
from several observing programmes that CO observations of nearby 
radio galaxies produce a high detection rate
(see e.g. Prandoni et al.~\cite{Pretal07}; Oca{\~n}a Flaquer et al.~\cite{Ocetal10}). Lim et al. (\cite{Letal00})
detected $^{12}$CO($1-0$) and $^{12}$CO($2 - 1$) 
emission from the FR\,I radio galaxies 3C\,31 and 264 with the IRAM 30m 
Telescope and established that the line profiles indicate disk rotation; this
appears to be a common property (Prandoni et al.~\cite{Pretal07}).

The derived masses of molecular gas are certainly adequate to power the jets in at
least some cases. For example, the well-studied FR\,I radio galaxy 3C\,31 has an estimated
jet power $P_{\rm jet} \approx 10^{44}$\,erg\,s$^{-1}$ (Laing \& Bridle~\cite{LB}), 
while the lifetime of the source must be $\geq 10^8/(v/0.01c)$ years, where $v$ is the 
advance velocity of the tails. 
This suggests that the lifetime of 3C31 may be appreciably longer than those of FR\,II sources 
($10^6-10^7$ years; Blundell \& Rawlings~\cite{BR00}) and therefore that a higher molecular gas
mass may be needed to fuel it, despite its lower radio luminosity. 
For an assumed conversion efficiency $\eta$ from the rate of 
gravitational energy release to jet power, the required mass
accretion rate for 3C 31 is $\dot{M} = P_{\rm jet}/\eta c^2 \approx 0.02M_{\odot}{\rm yr}^{-1}$ if $\eta = 0.1$ 
and the molecular gas mass needed to fuel the AGN for $10^8$ years would be 
$\sim 2\times 10^6 M\odot$ (assuming a constant accretion rate).
The mass of molecular gas associated with 3C\,31 is $\approx 10^9 
M_\odot$ (Lim et al.~\cite{Letal00}). This implies that, even if the efficiency is
significantly smaller or the advance speed is $<0.01c$, there is ample fuel for any reasonable source lifetime. 
3C\,31 is a
luminous FR\,I radio galaxy, and correspondingly smaller amounts of cold gas
would be required to fuel less powerful sources, assuming efficient accretion.
Current $H_2$ mass upper limits (of the order of $5\times 10^7-10^8 M\odot$; e.g. Fig. ~\ref{fig-H2massz}) 
do not rule out the possibility that cool gas is the dominant fuelling mechanism in 
low luminosity radio galaxies, as well as in their brighter counterparts.
If this is the case, we expect to detect CO in virtually all radio galaxies, provided that CO 
spectroscopy goes deep enough.
 
Only high-resolution imaging of CO or other molecular lines would allow us to probe 
the dynamics of the cool gas directly, to determine if and how much of 
it is infalling, and what kind 
of motions (rotation, non-circular, radial in- or out-flows) are present.  
Interferometry at mm wavelengths is challenging with current instrumentation, and 
has been possible for only a few objects. Interferometric observations of 3C\,31
by Okuda et al.~(\cite{Oetal05}) showed that the CO coincides spatially with 
the dust disk observed by HST (Martel et al.~\cite{Metal99}) and is in 
ordered rotation.  
These authors suggest that the cold gas is in stable orbits, in which case the
actual accretion rate may be very low.

ALMA, thanks to its combination of angular resolution and sensitivity, will give
us a unique opportunity to assess the cold accretion scenario in low luminosity
radio galaxies, by directly imaging the gas on scales $\sim$10 -- 100\,pc and
determining its dynamical state. Very little is known about cold gas in Southern
galaxies, however, with very few objects observed in CO.

In this paper we report the results of pilot APEX-1 CO line observations 
of Southern radio galaxies. 
In Sect.~2 the Southern radio galaxy sample is presented with particular 
emphasis on the three sources discussed in this paper. In Sect.~3
the APEX line measurements are described, while in Sect.~4  a brief 
discussion of the results is reported.
In the following we assume a standard $\Lambda$-CDM cosmology with $H_0=70$ km\,s$^{-1}$\,Mpc$^{-1}$.

\section{Sample Description}

\begin{figure}[t]
\vspace{0.5cm}
\resizebox{\hsize}{!}{\includegraphics[width=3.5cm]{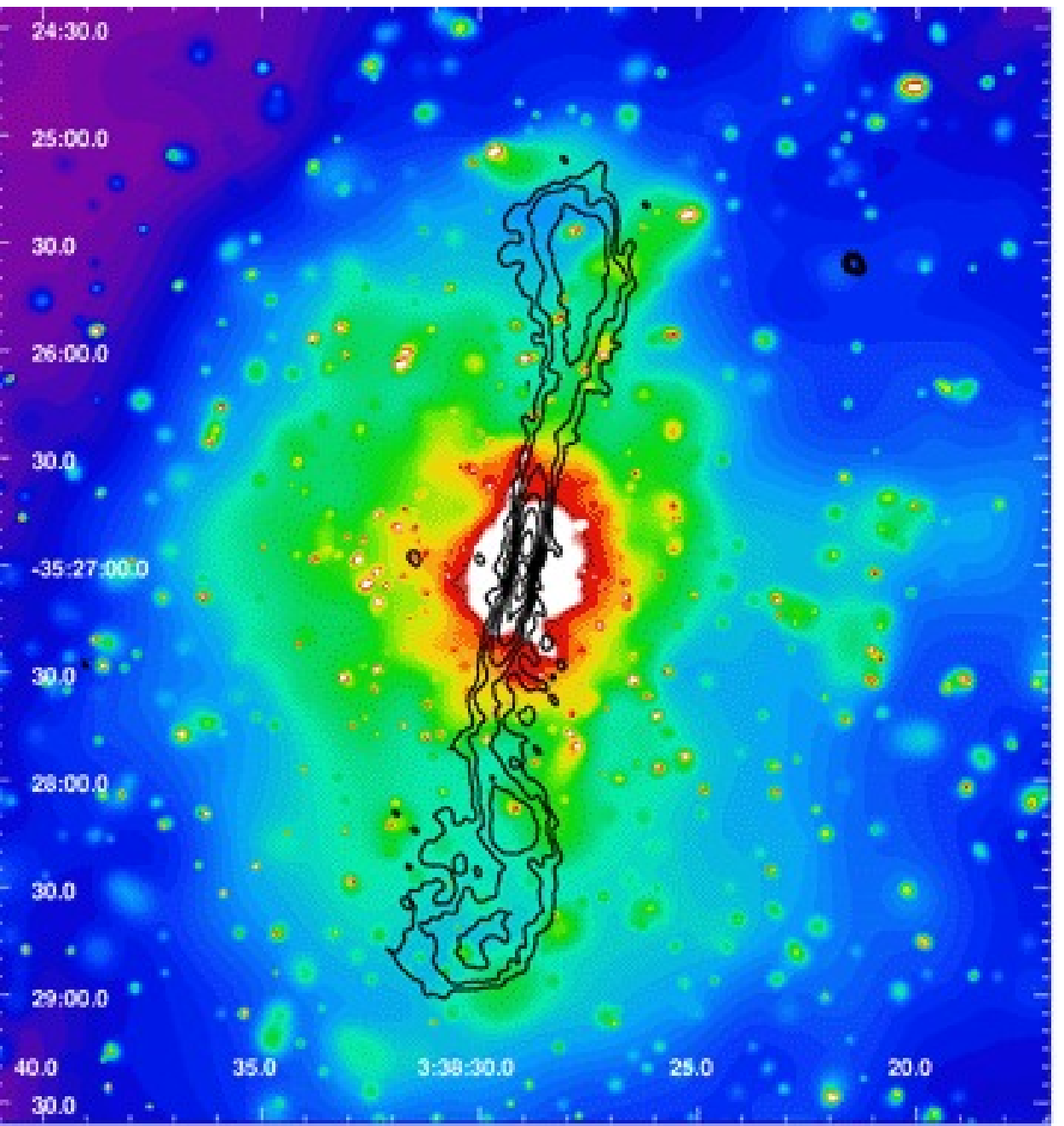}\hspace{0.5cm}\includegraphics[width=3.9cm]{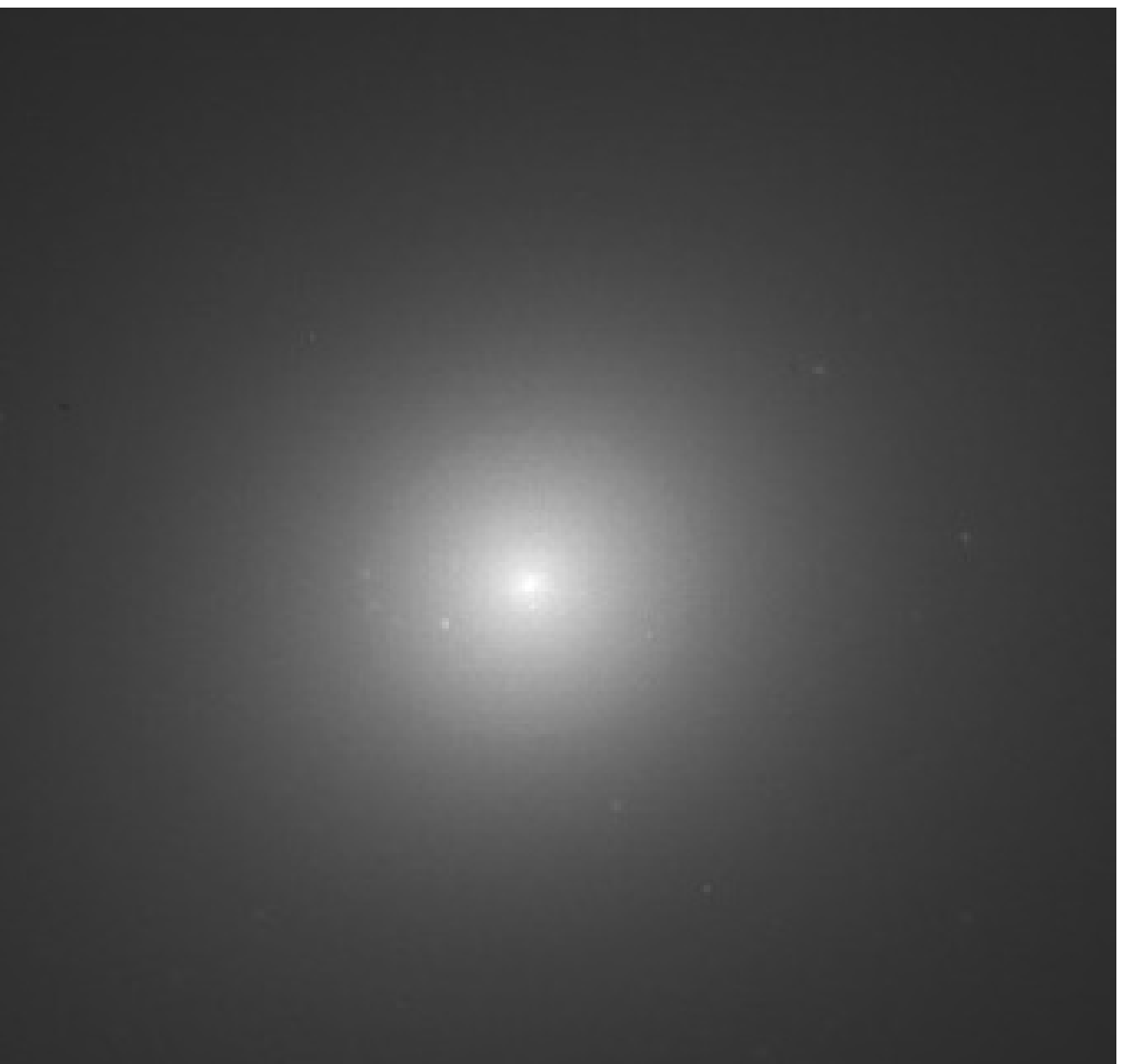}}

\vspace{1cm}

\resizebox{\hsize}{!}{\includegraphics[width=3.5cm]{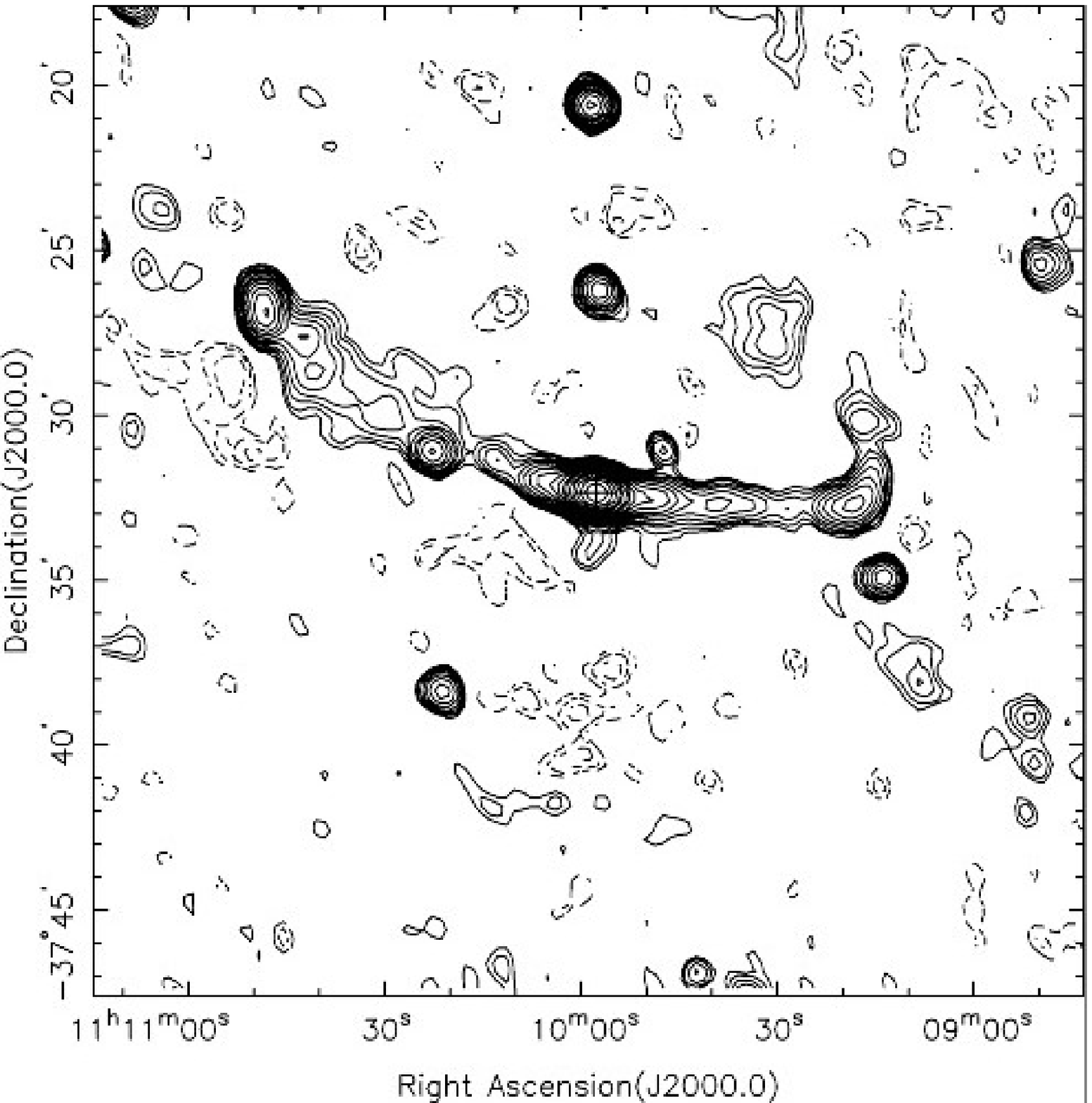}\hspace{0.5cm}\includegraphics[width=4.0cm]{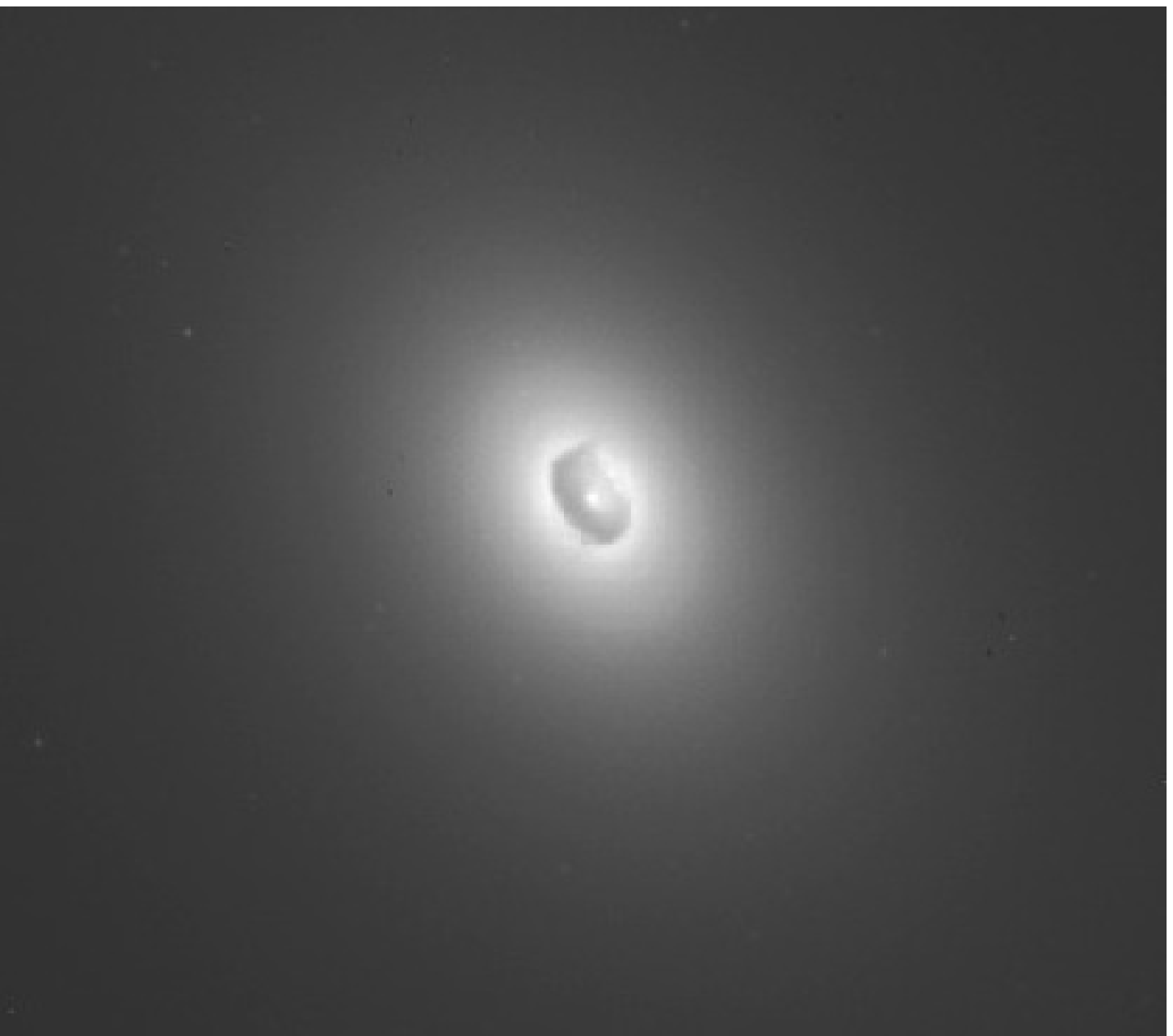}}

\vspace{1cm}

\resizebox{\hsize}{!}{\hspace{-1cm}\includegraphics[width=6.5cm]{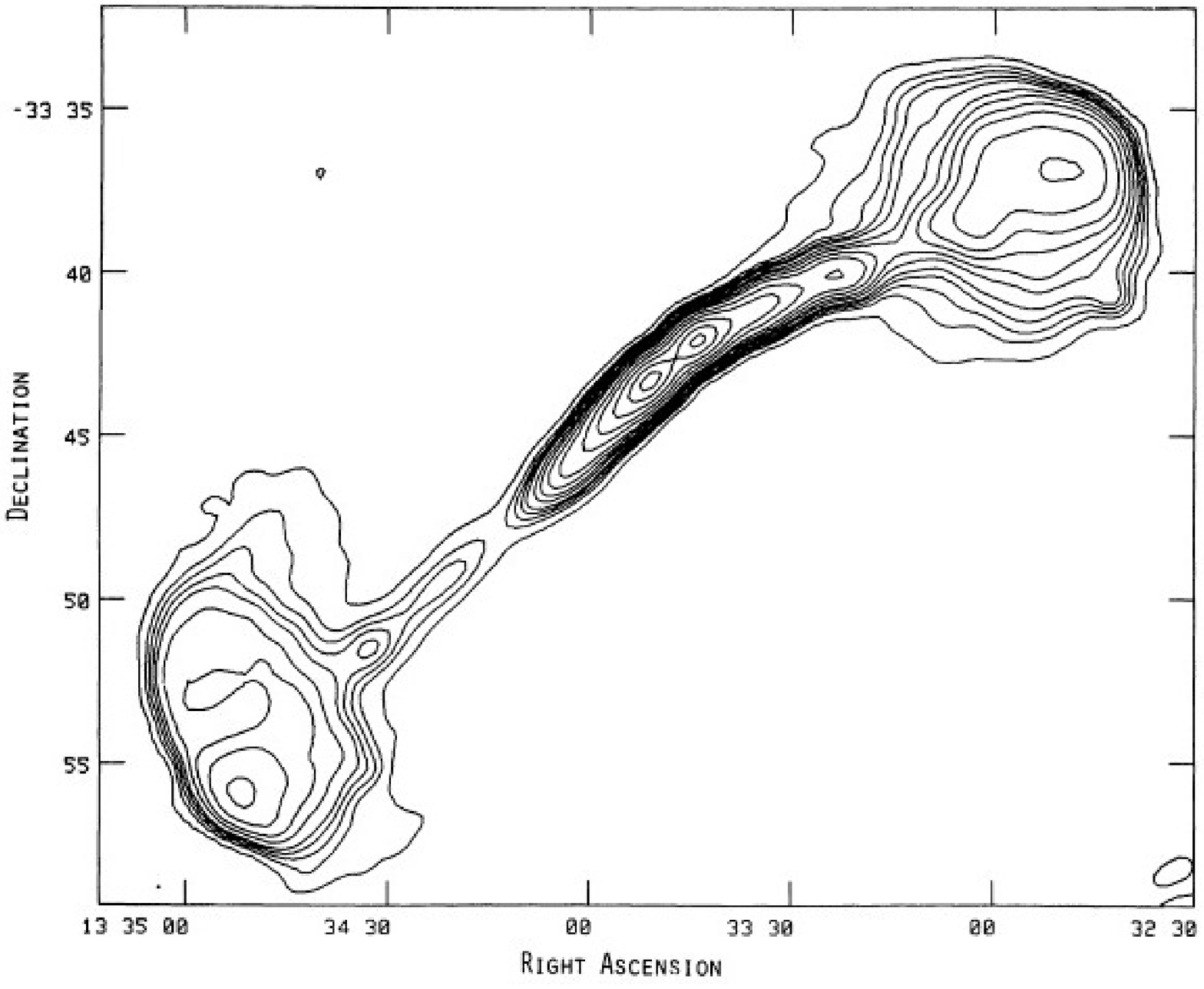}\includegraphics[width=5.5cm]{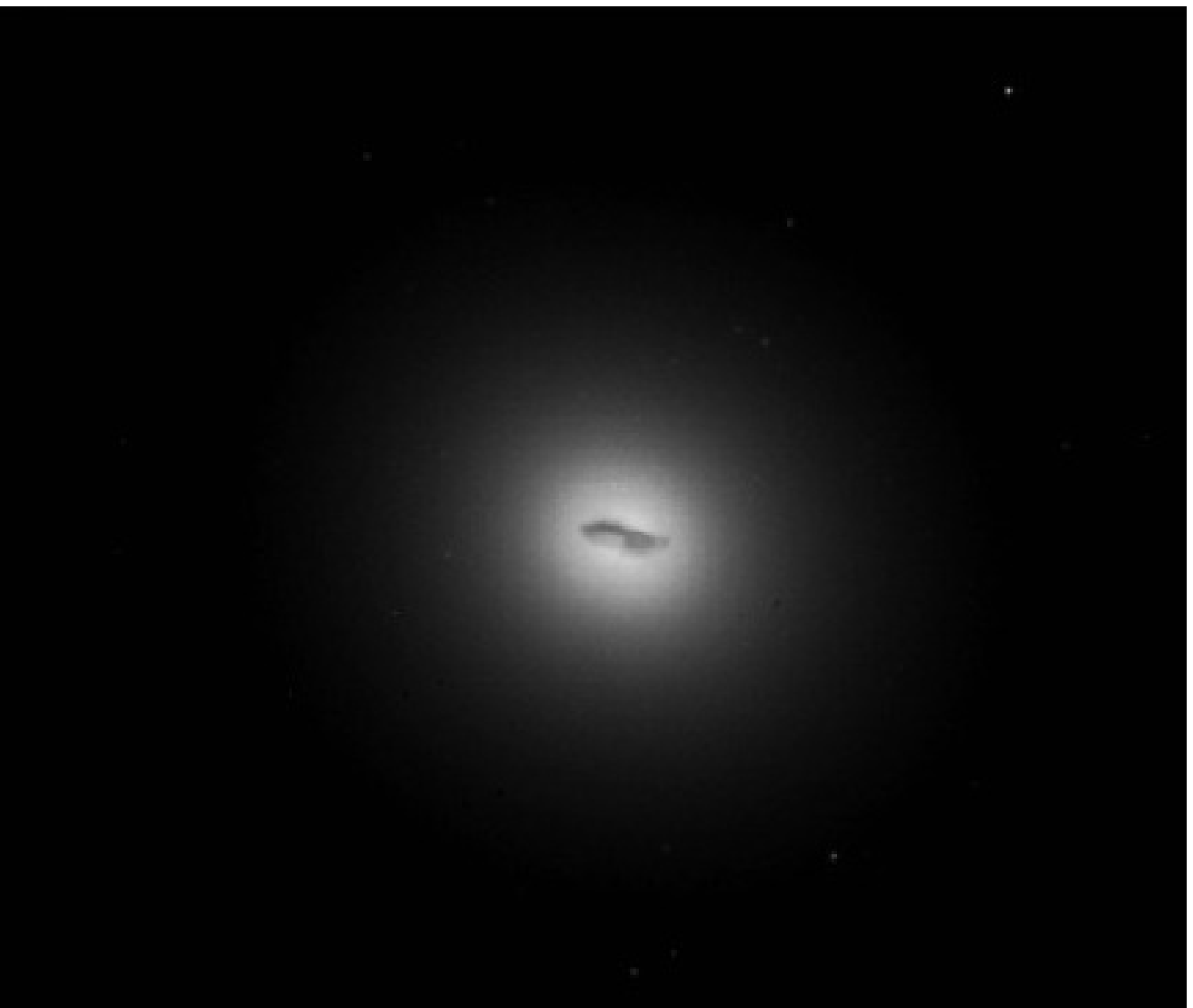}}

\vspace{1cm}

\caption[]{1.4 GHz VLA radio continuum contours ({\it left}) and HST images ({\it right}) for radio galaxies NGC\,1399 ({\it top}), NGC\,3557 ({\it middle}), IC\,4296 ({\it bottom}). 1.4 GHz radio continuum contour images are taken respectively from 
Shurkin et al. (\cite{Shetal08}) for NGC\,1399 (here overlaid to a Chandra X-ray color image); from the NVSS database (Condon et al.~\cite{Cetal98}) for NGC\,3557; and from Killeen et al. (\cite{Ketal86}) for IC\,4296. The HST images are taken from Lauer et al. (\cite{Lauetal05}).
\label{fig-hst}}
\end{figure}

\begin{table*}[t]
\caption{APEX-1 CO($2-1$) line measurements
\label{tab-meas}}
\centering
\begin{tabular}{llrrrccrcrr}
\hline\hline
\multicolumn{2}{l}{Source} &   
\multicolumn{1}{c}{$z$} & 
\multicolumn{1}{l}{$t_{ON}$} & 
\multicolumn{1}{l}{$T_{sys}$} & 
\multicolumn{1}{c}{$\Delta$v$_{\tt res}$} & 
\multicolumn{1}{l}{$T_{a}^{\rm rms}$} & 
\multicolumn{1}{c}{$\Sigma T_a d$v}  & 
\multicolumn{1}{c}{$\Delta$v$_{\tt FWHM}$} & 
\multicolumn{1}{c}{$\frac{T_a^{\rm peak}}{T_a^{\rm rms}}$} & 
\multicolumn{1}{c}{$\log{M_{H_2}}$}\\
\multicolumn{3}{c}{} &
\multicolumn{1}{c}{(min)} &
\multicolumn{1}{c}{(K)} &
\multicolumn{1}{c}{(km\,s$^{-1}$) } & 
\multicolumn{1}{c}{(mK)} &
\multicolumn{1}{c}{(K km\,s$^{-1}$) }  & 
\multicolumn{1}{c}{(km\,s$^{-1}$) } & 
\multicolumn{1}{c}{} & 
\multicolumn{1}{c}{(M$\odot$)} \\
\hline
PKS0336-35 & [\object{NGC\,1399}] & 0.0048 & 108.2 & 272  & 40 & 1 & 0.20 & 127 & 2.1 & 
7.47 \\
 &  &  &  &  & 120 & 0.35 & 0.39 & 365 & 3.3 & 7.75  \\
PKS1107-372 & [\object{NGC\,3557}] & 0.0102 & 18.4 & 316 & 40 & 2 & 1.52 & 248 & 3.5 & 
9.02 \\
PKS1333-33 & [\object{IC\,4296}] & 0.0125 & 37.4 & 315 & 40 & 1 & $<0.51$ & $\cdots$ & $\cdots$ & 
$<8.70$ \\
\hline
\end{tabular}
\vspace{0.5cm}
\end{table*}

The primary radio-galaxy samples (3C, B2) for which CO observations have 
been made are difficult to observe from the ALMA site. We have therefore built a volume 
limited sample of low luminosity radio galaxies in the 
Southern sky, for which we intend to obtain complete single-dish CO 
coverage in order to prepare for ALMA imaging.
This sample has been defined in a very similar way to the Northern sample we extracted 
from the B2 radio-source catalogue and followed up in CO with the 
IRAM 30m telescope (Prandoni et al.~\cite{Pretal07}).

The parent sample is selected from the Parkes 2.7-GHz survey, in the declination range  
$-17^{\rm o}<\delta < -40^{\rm o}$, as described by Ekers et al.\ (\cite{Eketal89}). 
It has a radio flux-density limit of 0.25 Jy 
at 2.7 GHz and an optical limit of $m_V \leq 17.0$, and consists of 191 radio 
galaxies. We extracted those sources with redshifts $z<0.03$ which are associated with elliptical galaxies.
The resulting 11 objects thus form a small, but volume-limited and 
complete subset of the Ekers et al. (\cite{Eketal89}) sample. 
All eleven sources of the volume-limited sample are of the FR\,I type.
The subset is quite similar to the sample of 3C and B2 radio galaxies in 
the Northern Hemisphere, which consists of 23 low luminosity radio sources   
associated with elliptical galaxies at $z<0.03$; 16 of these have been observed in CO with the IRAM 30m 
telescope (Prandoni et al.~\cite{Pretal07}).

As part of the science verification programme of the APEX-1 receiver, 
three of the eleven sources in our subsample were selected for 
observation in the $^{12}$CO($2-1$) line. 
They were: PKS0336-35 (\object{NGC\,1399}), PKS1107-372 (\object{NGC\,3557}), and PKS1333-33 
(\object{IC\,4296}). All three are radio sources with twin 
radio jets (see Fig.~\ref{fig-hst}, left panels), and had been observed with 
the WPFC2 of the Hubble Space Telescope (Lauer et al. \cite{Lauetal05}). 
While no trace of dust was found in the nucleus of NGC\, 1399, the 
other two show prominent dusty disks around their respective nuclei 
(see Fig.~\ref{fig-hst}, right panels).

A fourth sample member, PKS0320-37  (Fornax A) was imaged 
in $^{12}$CO($1-0$) and $^{12}$CO($2-1$) by Horellou et al.~(\cite{Horetal01}); it shows a complex
dust structure in HST observations (Grillmair et al.~\cite{Gretal99}).

No CO imaging is available for the remaining sample members.

\begin{figure*}[t]
\centering
\resizebox{\textwidth}{!}{\includegraphics[]{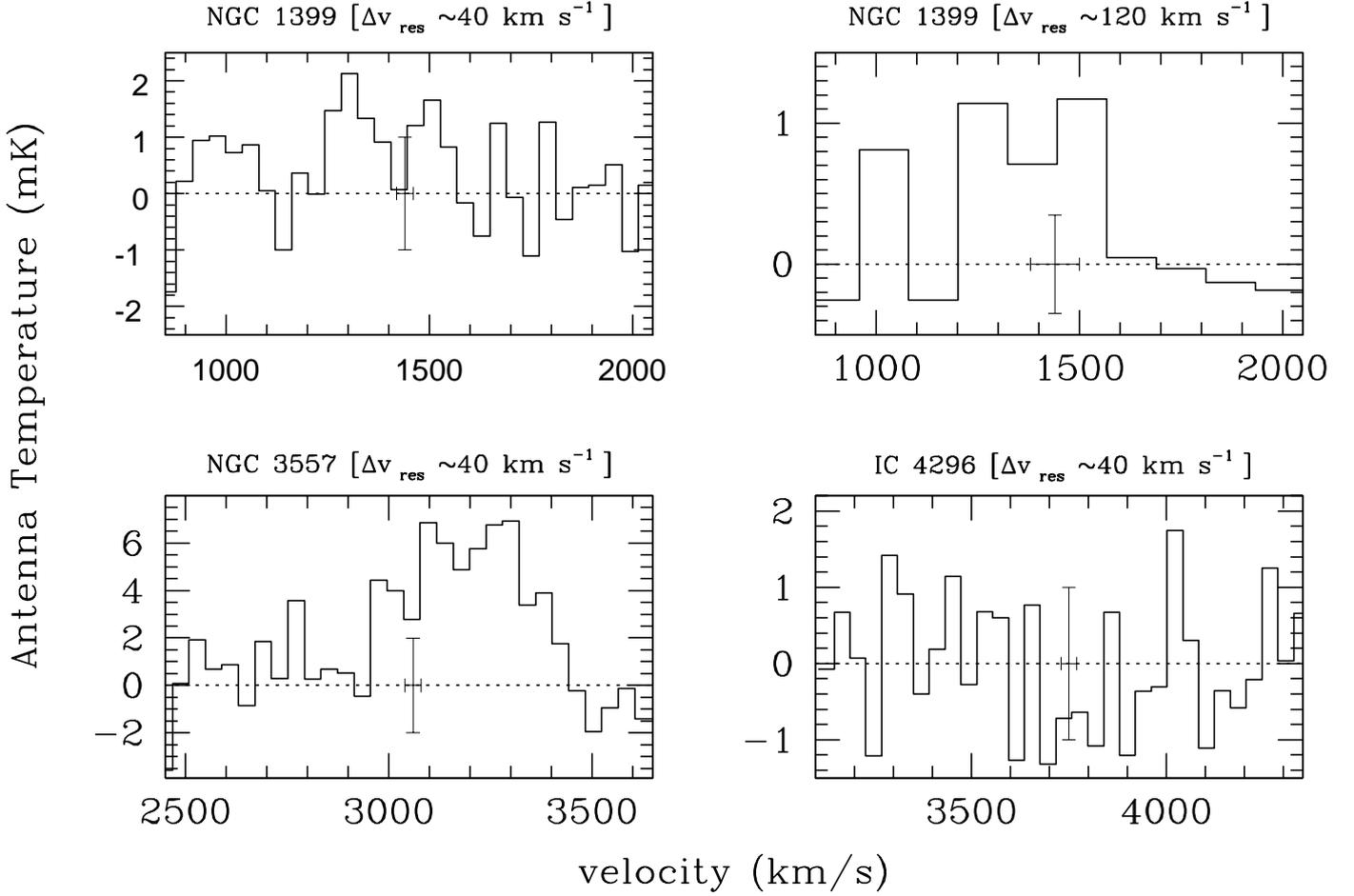}}
\vspace{-5.5cm}
\caption[]{$^{12}$CO($2-1$) spectra obtained at APEX 
for the radio galaxies NGC\,1399 (top), NGC\,3557 (bottom, left) and IC\,4296 (bottom, right). 
NGC\,1399 is shown for two different smoothing boxes: a velocity width of $\approx 40$ km\,s$^{-1}$ (top, left), as for the other two sources; and a wider velocity width of $\approx 120$ km\,s$^{-1}$ (top, right). Typical spectrum error bars ($\pm T_{a}^{rms}$ and $\pm \Delta$v$_{\tt res}/2$) are shown for reference at the systemic velocity of the galaxies. Horizontal dashed lines indicate the zero level $T_{a}=0$.
\label{co-lines}}
\end{figure*}

\section{Line Observations and Measurements} \label{obs}
We used the APEX-1 SHeFI receiver connected to the FFTS backend to search for emission in the 
(2-1) transition of $^{12}$CO.
Low luminosity radio galaxies have $^{12}$CO($2-1$) line widths 
typically 
spanning $200-600$ km\,s$^{-1}$  FWHM (Prandoni et al.~\cite{Pretal07}) and are therefore 
fully sampled by a 1-GHz wide individual unit of the FFTS backend 
(covering a total velocity range of 1300 km\,s$^{-1}$  at 230 GHz). 
The 30 arcsec FWHM beam 
probes the presence of molecular gas in the galaxy cores (the inner 3 -- 7 kpc, 
depending on the redshift), allowing a direct comparison with dust 
structures on similar scales imaged with HST.

The observations were carried out in Summer 2008 and the data were reduced with
the CLASS package. Scans severely affected by ripples and/or noisy baselines
were removed and first-order baselines were subtracted. The noise levels
obtained were typically $T_a^{\rm rms}\approx 1-2$ mK ($1\sigma$) at a reference smoothing width 
of $\Delta$v $\approx 40$ km\,s$^{-1}$. Line fluxes were measured by numerically
integrating over the channels in the line profile.  Line widths were measured as
full widths at half power.  A source was considered detected when the
$^{12}$CO($2-1$) emission line had $T_a^{\rm peak}>3 T_a^{\rm rms}$, and
tentatively detected when $T_a^{\rm peak}>2 T_a^{\rm rms}$.  In case of non
detections, upper limits were calculated, following Evans et
al. (\cite{Eetal05}), through the relation
\begin{equation}\label{eq-upper}
T_a \Delta \rm{v} \;\; (\rm{K \; km\,s^{-1} })\;\; < \frac{3 T_a^{rms} 
<\Delta \rm{v}>}{\sqrt{<\Delta \rm{v}>/\Delta \rm{v}_{res}}}  
\end{equation}
where $<\Delta \rm{v}>$ is the mean FWHM line width for radio galaxies (in our
case assumed equal to 500 km\,s$^{-1}$ , see Prandoni et al.~\cite{Pretal07}, Evans et al.~\cite{Eetal05}); 
$<\Delta \rm{v}_{res}>$ is the velocity resolution (in our case $\approx 40$ km\,s$^{-1}$  after smoothing) and 
\begin{equation}
T_a = T_{sys}/\sqrt{\Delta \nu \cdot t}
\end{equation}
where $\Delta \nu$ is the channel width ($\approx 30$ MHz, after smoothing) 
in Hz and $t$ is the on-source integration time in seconds.

H$_2$ molecular masses were derived using the same relation as used by Gordon et al. (\cite{GBC92}) 
and later by Lim et al. (\cite{Letal00}), but modified for a $\Lambda$-CDM cosmology and $H_0=70$ km\,s$^{-1}$\,Mpc$^{-1}$.
We assumed an aperture efficiency $\eta_a = 0.63$ for 230-GHz APEX observations.

A summary of our CO line measurements is given in 
Table~\ref{tab-meas}, where, for each source, we list the source redshift 
($z$); the on-source integration time ($t_{ON}$), after flagging data 
affected by poor baselines and/or ripples; the mean system temperature 
during the observations ($T_{sys}$); the channel width after smoothing ($\Delta$v$_{\tt res}$) and the corresponding noise 
level in the line spectra ($T_a^{\rm rms}$); the line integrated flux ($\Sigma T_a d\rm{v}$); the
FWHM line width ($\Delta\rm{v}_{\tt FWHM} $); the line signal to noise 
($T_a^{peak}/T_a^{rms}$) and the molecular gas mass ($\log{M_{H_2}}$). The final 
CO line spectra are shown in Fig.~\ref{co-lines}. 

NGC\,3557 was easily detected at a channel resolution of $\Delta$v$_{\tt res}\approx 40$ km s$^{-1}$, while for IC\,4296 
(the source at highest redshift) only an upper limit could be derived. In the case of NGC\,1399 only the higher of the two peaks in
the spectrum (Fig.~\ref{co-lines}, top left) has a signal-to-noise ratio $T_a^{peak}/T_a^{rms}>2$ and can be considered tentatively 
detected. However the systemic velocity of the galaxy falls between the two peak. By increasing the smoothing to 
$\Delta$v$_{\tt res}\approx 120$ km s$^{-1}$ (Fig.~\ref{co-lines}, top right), we can reduce  
the noise level to $T_a^{rms}\sim 0.35$ mK and get a clear $>3\sigma$ detection over a wider velocity range. In this case we obtain 
$\Sigma T_a d\rm{v}=0.39$  K km s$^{-1}$, $\Delta \rm{v}_{\tt FWHM} = 365$ km\,s$^{-1}$, and $\log{M_{H_2}/M\odot}=7.75$ 
(see Table~\ref{tab-meas}). 
NGC\,1399 does not show detectable dust features in the HST image, and the presence of CO is particularly interesting in this case.
There is some evidence for a double-horned structure in the spectrum of NGC\,3557, although the statistical significance is low 
(Fig.~\ref{co-lines}, bottom left). If confirmed, the presence of CO in ordered rotation would be consistent with the strong
dusty disk seen in HST observations (see Fig.~\ref{fig-hst}). 
A very marked dusty disk is present even in IC\.4296 (see Fig.~\ref{fig-hst}) and we suspect that a longer observation ($t_{ON} \gg 30$ min), possibly with better weather conditions ($T_{sys} < 300$ K), will lead to a detection.

\section{Discussion and Conclusions}\label{sec-comp}

\begin{figure}[t]
\resizebox{\hsize}{!}{\includegraphics[]{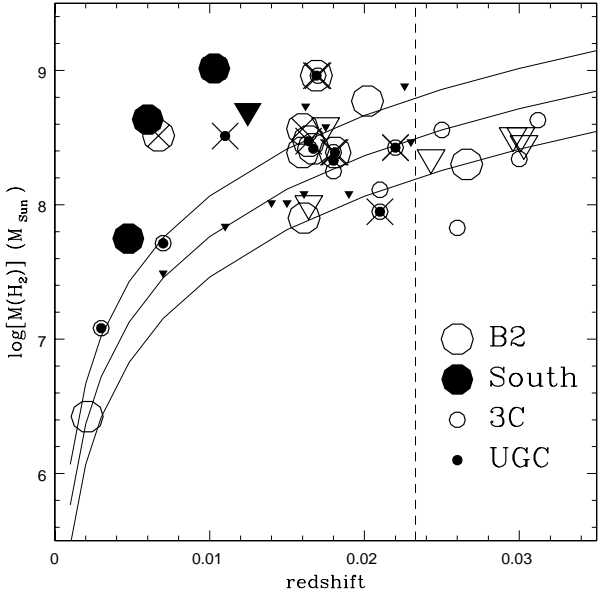}}
\caption[]{$H_2$ mass vs redshift for the 
$z<0.03$ B2 radio galaxies (large blue empty symbols); 
for $z<0.03$ 3C radio galaxies (small green empty symbols); for 
$z<0.0233$ UGC 
galaxies (small red filled symbols); and for the four (including Fornax A) 
$z<0.03$ Southern radio galaxies observed so far (large magenta filled symbols).
Circles refer to detections and triangles to upper limits. 
Crosses indicate sources which display a double-horned CO line profile.
Solid lines indicate  the $M_{H_2}$ 3$\sigma$ detection limit for
$\sigma$ ($=T_{\rm a}^{\rm rms}$) =  0.1, 0.2 and 0.4 mK respectively (assuming a 500 km\,s$^{-1}$  wide 
CO line, see equation~\ref{eq-upper}). The vertical dashed line at $z=0.0233$ 
indicates the redshift limit to which all three Northern samples overlap. 
\label{fig-H2massz}}
\end{figure}

The H$_2$ masses of the four low-luminosity southern radio galaxies observed so
far (including Fornax A) are plotted against redshift as magenta symbols in
Fig.~\ref{fig-H2massz}, together with the 23 B2 $z<0.03$ radio galaxies (blue
symbols, Prandoni et al.~\cite{Pretal07}), the brighter $z<0.03$ 3C radio galaxies studied
by Lim et al. (\cite{Letal03}; green symbols) and the $z<0.0233$ UGC galaxies
with radio jets studied by Leon et al. (\cite{Leonetal03}; red symbols). We note
that the three northern samples are partially overlapping and that some upper
limits have have been updated using recent measurements by Oca{\~n}a Flacquer et
al. (\cite{Ocetal10}).  The deepest CO line observations with existing
facilities have $3\sigma$ detection limits $\approx 3\times 0.1$\,mK, as indicated by
the solid lines in Fig.~\ref{fig-H2massz}. CO detection rates vary between 45\%
and 65\% for the three samples, but these differences are not significant once
account is taken of the relative sensitivities and redshift distributions. We
find that molecular gas masses are very similar over the full range of redshift
(and, implicitly, radio power) for the combined samples, spanning the range
$\approx 10^7 - 10^9$ M$_\odot$, with upper limits varying from $\approx 10^7$
M$_\odot$ to $\approx 10^8$ M$_\odot$, depending on distance. These values are
consistent with the hypothesis that the jets of low-luminosity (FR\,I) radio
galaxies can be powered by accretion of cold gas, but there is a large scatter
in molecular gas mass at a given radio luminosity, as also found by McNamara et
al. (\cite{McN2010}) for brightest cluster galaxies.

Seven examples of CO detections without visible dust, including NGC\,1399, are plotted in  Fig.~\ref{fig-H2massz}. 
The inferred $H_2$ masses range from $8\times 10^7$ to $4\times 10^8$ $M\odot$ (Oca{\~n}a Flaquer et al.~\cite{Ocetal10}; 
this paper). It is therefore not yet clear whether there is a significant  difference between the $H_2$ masses of radio 
galaxies with and without dust.
Nevertheless CO line detections are frequent in radio galaxies with dust, even more when the dust is in form of nuclear disks. 
In the sample of 23 elliptical B2 radio galaxies with $z<0.03$ (Prandoni et al.~\cite{Pretal07}), seventeen objects 
have detailed optical imaging (mostly with HST), and ten show clear dusty disks. CO line 
observations are available for nine of these objects, and $\approx 80\%$ (7/9) were detected. 
We also find that almost one half of the objects with dusty disks detected in CO have double-horned line profiles. 
The double-horned CO line profiles characteristic of ordered rotation (indicated
by the crosses in Fig.\ref{fig-H2massz}) occur over a wide range of molecular
gas masses.  Since many of the other detections are still marginal, it is possible that the
molecular gas in the majority of low-luminosity radio galaxies is in the form of
kpc-scale nuclear disks or rings, cospatial with the dust and in ordered rotation.  
There are exceptions, however: Fornax A represents a more complex
case where the CO shows a multi-component line profile associated with the complex
dust structures visible in HST $B-I$ images. The most probable scenario is that
the gas comes from the accretion of one or more small gas-rich galaxies
(Horellou et al.~\cite{Horetal01}).
The fact that we are finding CO in radio galaxies with no detectable dust 
may suggest that in (at least some) objects the bulk of the gas is in a cold (molecular) phase, supporting 
the idea that virtually all radio galaxies may be fuelled by cold gas. 

These results raise a number of questions which can be answered by observations
at optical, mid-infrared and mm wavelengths, and in particular by 
sensitive, high-resolution imaging with ALMA, as follows.
\begin{enumerate}
\item Is the molecular gas in disks or in rings with well-defined inner edges?
\item Is the majority of the gas in stable orbits?
\item What is the evidence for non-circular motions, especially infall?
\item Where does the gas come from: is its angular momentum consistent with
  that of the stellar population or different, indicating an external origin?
\item Do the molecular and ionized gas distributions have the same spatial
  distributions and kinematics? Does the gas become progressively hotter and
  then ionized closer to the nucleus?
\item Is there evidence for ongoing or recently-completed star formation?
\end{enumerate}

In conclusion, we have shown that observations with APEX in the 230-GHz band can
efficiently detect CO in low-luminosity radio galaxies. We plan to use APEX to
obtain CO spectroscopy for the entire southern sample and to build up a target list for 
imaging with ALMA.

\begin{acknowledgements}
The authors thank the APEX team, which successfully overcame a number of
technical obstacles to obtain the APEX-1 science verification observations
presented here. The authors would like to thank the referee, whose suggestions 
significantly improved the quality of the paper.
\end{acknowledgements}

\bibliographystyle{aa}

\end{document}